\documentclass[onecolumn,showpacs,preprintnumbers,amsmath,amssymb]{revtex4}
\usepackage{graphicx}
\usepackage{graphics}
\usepackage{dcolumn}
\usepackage{bm}

\def\BEq{\begin{equation}}
\def\EEq{\end{equation}}
\def\BEqA{\begin{eqnarray}}
\def\EEqA{\end{eqnarray}}
\def\BEn{\begin{enumerate}}
\def\EEn{\end{enumerate}}
\def\BWT{\begin{widetext}}
\def\EWT{\end{widetext}}

\def\bra{\langle}

\def\ket{\rangle}

\begin{document}


\title{Greenberger-Horne-Zeilinger state protocols for fully connected qubit networks}
\author{Andrei Galiautdinov}
 \email{ag@physast.uga.edu}
\affiliation{
Department of Physics and Astronomy,
University of Georgia, Athens, GA 30602}
\author{Mark W. Coffey}
\email{mcoffey@mines.edu}
\author{Ron Deiotte}
\affiliation{Department of Physics, Colorado School of Mines, Golden, CO 80401}

\date{\today}

\begin{abstract}

We generalize the recently proposed Greenberger-Horne-Zeilinger (GHZ) tripartite protocol 
[A. Galiautdinov, J. M. Martinis, Phys. Rev. A {\bf 78},  010305(R) (2008)] to fully 
connected networks of weakly coupled qubits interacting by way of anisotropic Heisenberg
exchange $g\left(XX+YY\right)+\tilde{g}ZZ$. Our model adopted here differs from the more familiar 
Ising-Heisenberg chain in that here every qubit interacts with every other qubit in the circuit. 
The assumption of identical couplings on all qubit pairs allows an elegant proof of the protocol for 
arbitrary $N$. In order to further make contact with experiment, we study fidelity degradation due to 
coupling imperfections by numerically simulating the $N=3$ and $N=4$ cases. Our simulations indicate 
that the best fidelity at unequal couplings is achieved when (a) the system is initially 
prepared in the uniform superposition state (similarly to how it is done in the ideal case), and (b) 
the entangling time and the final rotations on each of the qubits are appropriately adjusted.

\end{abstract}

\pacs{75.10.Jm, 03.67.Bg, 03.67.Lx, 85.25.-j}    

\maketitle



\section{Introduction}

In the rapidly developing field of quantum computing the concept of quantum entanglement 
is considered to be of greatest practical importance. Many important applications, such as quantum
communication \cite{PERESTERNO}, secret sharing \cite{SECRETSHARING}, open-destination teleportation 
\cite{OPENDESTINATIONTELEPORTATION}, fault-tolerant 
computing \cite{FTQCSHOR, FTQCKNILL}, and others \cite{NIELSEN}, rely heavily on an architecture's ability 
to generate multipartite entangling states. 
This is especially true for the general Greenberger-Horne-Zeilinger (GHZ) states \cite{GHZoriginal, GHZwithSH}
\BEq
|{\rm{GHZ}}\ket_N = {1 \over \sqrt{2}}(|0\ket^{\otimes N} + |1\ket^{\otimes N}),
\EEq
also known as cat states (see, e.g., \cite{LEIBFRIED2005}),
which, according to most entanglement measures, are {\it maximally} entangled \cite{NIELSEN}.

Various approaches to generating GHZ states have been proposed in the literature.
A scheme for generating cat states of a single-mode optical field by means of conditional 
measurement was proposed in Ref. \cite{DAKNA97}. A one-step multi-atomic GHZ state generation 
in a non-resonant cavity by way of cavity-assisted collisions was considered in Ref. \cite{ZHENG01}.  
Ref. \cite{BODOKY2007} discusses the possibility of generating GHZ states of electron spin qubits in a chain 
of quantum dots using the naturally available single-qubit rotations and the two-qubit Heisenberg exchange 
interaction. The minimum number of required operations in that proposal scales linearly with the number of 
qubits. In Ref. \cite{MATSUO07}, in the context of capacitively coupled superconducting phase qubits, an 
$N$-qubit GHZ state is proposed to be generated by applying an initial one-qubit $\pi/2$-pulse followed by 
an alternating sequence of $2(N-2)+1$ controlled-NOT (CNOT) and SWAP gates.  
In Ref. \cite{CIRAC08} it is shown how the GHZ state can be generated in a multi-qubit system by coupling 
it sequentially to a mesoscopic static completely mixed spin bath.
In Ref. \cite{BISHOP09}, in the context of circuit quantum electrodynamics, high-fidelity GHZ states 
(in a system whose mutual qubit detunings are large compared to the inter-qubit coupling strengths) are 
proposed to be probabilistically generated using one-qubit rotations and a single $N$-qubit dispersive readout.

In this paper we describe how GHZ states can be generated using a single $N$-qubit entangling pulse in a 
fully connected network of qubits (see also \cite{ASHHABfullyConnected}) interacting by way of anisotropic 
Heisenberg exchange \cite{maxEnt1}
\BEq
\label{eq:H}
H_{\rm int} =
 \frac{1}{2}{\sum^{N-1}_{\ell= 1}}\sum^N_{k=\ell+1}
 \left[g
 \left(
 \sigma_x^\ell\sigma_x^k+\sigma_y^\ell\sigma_y^k \right)
 +\tilde{g}\sigma_z^\ell\sigma_z^k\right],
\EEq
where $\sigma_\mu^\ell$, $\mu = x,y,z$, are the Pauli
matrices $\sigma_\mu$ acting on the $\ell$-th qubit.
We concentrate on the case $g\neq \tilde{g}$ only, which
in the context of Josephson phase qubits (our initial motivation \cite{MARTINIS2006}) corresponds to 
either capacitive ($\tilde{g}=0$) or inductive ($0< |\tilde{g}/g|<0.1$) coupling scheme. 
We show that the duration of the entangling pulse is given by
\BEq
t_{\rm GHZ} = \pi/2|g-\tilde{g}|,
\EEq
and thus in a typical experiment with $g \sim 10$ MHz the time $t_{\rm GHZ}$ is on the order of $25$ ns,
which is small compared to the usually achieved coherence times of $\sim 500$ ns
\cite{MARTINIS2009}.
We point out that the interaction Hamiltonian given in Eq. (\ref{eq:H}) is the result of first projecting the 
exact system Hamiltonian onto the qubit subspace, switching to the interaction picture, and then 
applying the rotating wave approximation (RWA) in which the fast oscillating terms are ignored. 
Here, the time scale is set by a typical time to do a qubit operation, such as, e.g., the entangling 
time given above. The fast oscillations are the ones that occur at qubit transition frequencies, 
usually on the order of 10 GHz, and thus having transition times on the order of 0.1 ns \cite{maxEnt1}.

Why employ fully connected networks?

If we adopt an important simplifying assumption of identical couplings on all qubit pairs, then 
the answer
to the above question is: ``symmetry, simplicity, and experimental practicality.'' As will be shown 
below, the symmetric coupling 
scheme allows a rigorous proof of the protocol for arbitrary $N$. The resulting GHZ sequence 
is, indeed, very simple: prepare the system in the uniform superposition state, turn the coupling 
on and wait for a given 
time $t_{\rm GHZ}$, then apply a final, corrective $N$-qubit rotation. 

There is a nice experimental realization of fully connected networks involving capacitively 
coupled superconducting qubits proposed by Matthew Neeley and his collaborators at 
UCSB \cite{NEELEY}. It consists of 
a superconducting island (an isolated piece of aluminum) 
connected via a capacitor to each of the qubits in the system.
Using the standard methods of circuit analysis, one can transform
this ``star'' configuration into an equivalent ``delta'' (or, fully connected network) configuration,
in which there is a mutual capacitance between each pair of qubits. 
[The term "delta" comes from
the three-qubit case, where the system looks like a triangle, a special case of
the ``star-delta'' or ``Y-delta'' transform in electrical engineering \cite{LORRAIN}.]
It is then straightforward to show that $N$ qubits coupled to an island with
capacitance $C_{\star}$ are equivalent to $N$ qubits coupled to each other in
a complete graph with capacitance $C_\Delta = C_{\star}/N$ \cite{NEELEY}. 
Thus, from the practical point of view, implementing a fully connected network using a solid state quantum 
computing architecture is relatively straightforward. There are still some residual errors in the coupling that 
are due to 
imperfections in the coupling capacitors, so the resulting pairwise couplings are not strictly 
identical. That somewhat limits applicability of the model and, in a rigorous analysis, must be taken into 
account. Below we show how the errors due to imperfections in the individual couplings can be drastically 
reduced by adjusting the entangling time and the final rotations in the $N=3,4$ cases.

Our final motivation for the use of fully connected networks comes from their applicability to problems in 
number theory. Two of us (MWC and RD) have recently proposed an effective way to perform 
primality testing using this model with a modest number of qubits \cite{MC&RD}.

We now turn to the description and the proof of the GHZ protocol itself.

\section{GHZ protocol}
\label{sec:GHZprotocol}

First, notice that for any number of qubits $N$ the Hamiltonian (\ref{eq:H}) has eigenenergy 
(see Appendix \ref{sec:appendix1})
\BEq
E_{\rm GHZ} \equiv
\lambda_0(N) = C^2_N \left(\tilde{g}/2\right)
\EEq
associated with the eigenstates 
$|0\ket^{\otimes N}$ and $|1\ket^{\otimes N}$, where $C^2_N = (N-1)N/2$ is the total number of pairwise 
couplings present in the network.
Then, the two families of the GHZ protocols are
\BEq
N = 2,4,6, ... : \quad
e^{-i\lambda_0 t_{\rm GHZ}} e^{i \left(N/2-1\right) \pi}
|{\rm GHZ}\ket_{N} = R_3R_1U_{\rm ent}R_1 |0\ket^{\otimes N},
\EEq
and
\BEq
N=3,5,7, ... : \quad
e^{-i\lambda_0 t_{\rm GHZ}} e^{i(-1)^{(N-3)/2} (\pi/4)}
|{\rm GHZ}\ket_{N} = R_2 U_{\rm ent}R_1 |0\ket^{\otimes N},
\EEq
where
\BEq
R_3 = e^{-i \left[2+(-1)^{N/2}\right]\pi \sigma^1_z/4}, \quad
R_1 =  \otimes_{k=1}^N e^{-i\pi \sigma^k_y/4}, \quad
R_2 = \otimes_{k=1}^N e^{-i\pi \sigma^k_x/4},
\EEq
are the corresponding local rotations, and
$U_{\rm ent} = e^{-iH_{\rm int}t_{\rm GHZ}}$
is the entangling pulse of duration $t_{\rm GHZ} = \pi/2|g-\tilde{g}| $.

To see in detail how these protocols work we first re-write
the {\it fully} uniform superposition state,
\BEq
|\psi\ket_{\rm uniform} = R_1 |0\ket^{\otimes N} \equiv \frac{1}{2^{N/2}}\sum_{\xi_k\in \{0,1\}}|\xi_1\ket 
\dots |\xi_N\ket ,
\EEq
as a sum of 
{\it partial} uniform superpositions, each of which is characterized by the total number $j$ of up spins in the 
respective direct-product components, as follows:
\BEq
|\psi\ket_{\rm uniform} = \frac{1}{2^{N/2}} \sum_{j=0}^{N} \sqrt{C^j_N}\; |W_j\ket.
\EEq 
Here, 
$|W_j\ket$ stands for a generalized $W$-state (with $j$ spins up), and
$C^j_N = N!/j!(N-j)!$ is the corresponding binomial expansion coefficient. For example, 
for $N=3$, $|W_{1}\ket = [|001\ket+|010\ket + |100\ket]/\sqrt{3}$, 
$|W_{2}\ket = [|011\ket+|101\ket + |110\ket]/\sqrt{3}$, etc. 
Notice that for any $N$, 
\BEq
[|W_0\ket + |W_N\ket ]/\sqrt{2} \equiv [|0\ket^{\otimes N}+|1\ket^{\otimes N}]/\sqrt{2}
=  |{\rm GHZ}\ket_{N}. 
\EEq

We now state an important property of the interaction Hamiltonian $H_{\rm int}$:

\noindent {\bf The eigenvalue formula:} 
For any $j=0,1,2,\dots, N$, the states $|W_j\ket$ are the eigenstates of $H_{\rm int}$ associated with 
the eigenvalue 
\BEq
\label{eq:evalues}
\lambda_j(N) = j(N-j)(g-\tilde{g}) + C^2_{N}(\tilde g/2).
\EEq 
{\bf Proof outline.} The operator $\Sigma_z =\sum_k \sigma_z^k$ commutes with $H_{\rm int}$, providing a good 
quantum number, the total number of up spins in a given state.
Since $H_{\rm int}$ and $\Sigma_z$ share eigenvectors, the $|W_j\ket$ 
are eigensubspaces. 
The proof of the analytical formula (\ref{eq:evalues}) for the eigenvalues
is given in Appendix \ref{sec:appendix1}. Notice that since the Hamiltonian (\ref{eq:H}) commutes with 
the operator $X^{\otimes N}
=\mbox{NOT}^{\otimes N}$, if $|W_j\ket$ is an eigenvector, then so is $|W_{N-j}\ket$,
with the same eigenvalue. This is a powerful property, as it immediately tells us that the 
eigenenergies are constrained to satisfy the relation $\lambda_j = \lambda_{N-j}$.
$\blacksquare$

The eigenvalue formula immediately allows us to write down the effect of the entangling pulse on the uniform 
superposition,
\BEq
\label{eq:UentPsiuniform}
U_{\rm ent}|\psi\ket_{\rm uniform} = \frac{1}{2^{N/2}} \sum_{j=0}^{N}
 \sqrt{C^j_N}\;e^{-i\lambda_j t_{\rm GHZ}} |W_j\ket ,
\EEq
giving the state on which the final $N$-qubit rotations $R_3R_1$, $R_2$ will be acting. 
Assuming $g>\tilde{g}$, we have the useful relation,
\BEq
\label{eq:1}
e^{-i\lambda_j t_{\rm GHZ}}
=e^{-i\lambda_0  t_{\rm GHZ}}\exp\{-i[j(N-j)(g-\tilde{g})t_{\rm GHZ}]\}
=e^{-i\lambda_0  t_{\rm GHZ}}\exp[-i(\pi/2)j(N-j)]
=e^{-i\lambda_0  t_{\rm GHZ}}(-i)^{j(N-j)}.
\EEq

Now, for $N$ odd (Appendix \ref{sec:appendix2}), 
\BEq
\label{eq:2}
R_2^{-1} |{\rm GHZ}\ket_{{\rm {N ~odd}}}
={1 \over 2^{N/2}}\sum_{j=0}^N \sqrt{C_N^j} \frac{i^j [1+(-1)^j i^N]}{\sqrt{2}}|W_j\ket.
\EEq
In this case $i^N =\pm i$. Recalling that 
\BEq
\bra W_j|W_k\ket = \delta_{jk},
\EEq
we see that the full state is properly normalized. 
Combining Eqs. (\ref{eq:UentPsiuniform}),
(\ref{eq:1}), (\ref{eq:2}), with the identity
\BEqA
i^{j(N-j)}&=&(-i)^{j(N-j)} =\cos[j(N-j)\pi/2] \nonumber \\
&=&\exp\left[i(-1)^{(N+1)/2}{\pi \over 4}\right]{i^j \over \sqrt{2}}[1+(-1)^j i^N]
=\exp\left[i(-1)^{(N-3)/2}{\pi \over 4}\right]{i^j [1+(-1)^j i^N] \over \sqrt{2}},
\EEqA
gives the GHZ state protocol for odd $N$.

For $N$ even (Appendix \ref{sec:appendix3}), 
\BEq
\label{eq:3}
R_1^{-1} [R_{3}]^{-1}|{{\rm {GHZ}}}\ket_{{\rm {N ~even}}}
={1 \over {2^{N/2}}}\sum_{j=0}^N \sqrt{C_N^j} \frac{(-1)^j + e^{-i\theta}}{\sqrt{2}}|{W_j}\ket,
\EEq
where 
\BEq
\label{eq:theta}
\theta(N)=(\pi/2)[2+ (-1)^{N/2}].
\EEq
Combining Eqs. (\ref{eq:UentPsiuniform}) and
(\ref{eq:3}) with the identity
\BEq
(-i)^{j(N-j)}=e^{i\pi(N/2-1)} {e^{i\theta/2} \over \sqrt{2}}[(-1)^j +e^{-i\theta}],
\EEq
gives the GHZ state protocol for even $N$. [When $g<\tilde{g}$, and $N$ is even,
an extra factor of $e^{-i(\pi/4)(\sigma^2_z+\sigma^3_z)}$ should be inserted after $R_3$, otherwise 
the GHZ state accumulates an internal phase of (-1).]

Notice that for $g={\tilde g}$, the uniform superposition state is an eigenstate of the interaction Hamiltonian 
$H_{\rm int}$. Thus, the $|\psi\ket_{\rm uniform}$ does not change (apart from accumulating an overall phase) 
under the action of the entangling pulse $U_{\rm ent}$. This explains the previously stated requirement 
$g\neq {\tilde g}$ (cf. \cite{maxEnt1}).

\section{Fidelity degradation due to coupling errors}

\begin{figure*}
\includegraphics[angle=0,width=1.00\linewidth]{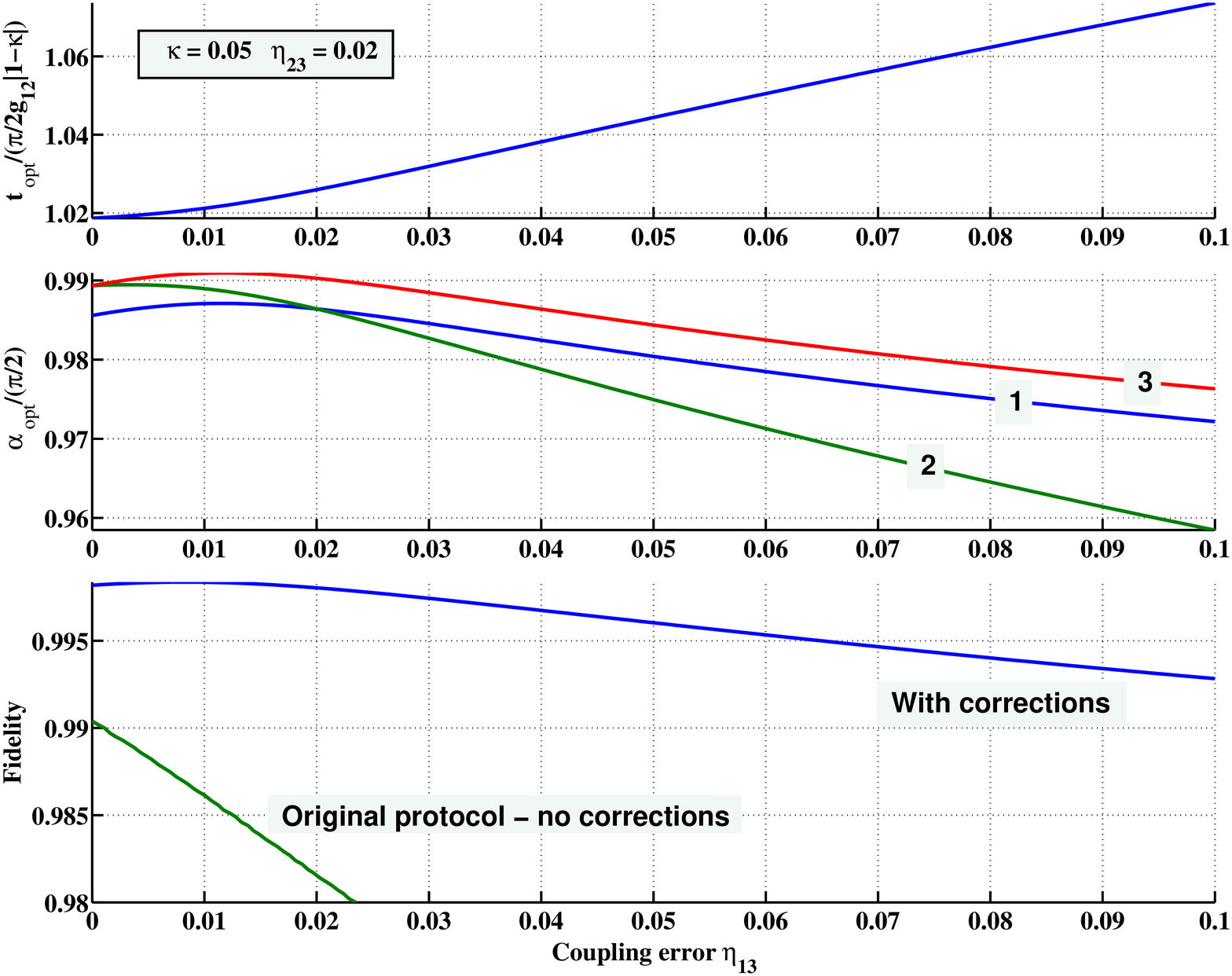}
\caption{
\label{fig:1} 
(Color online) Optimal values of the final rotation angles $\alpha_{\rm opt}^i$, $i=1,2,3$, and the entangling time 
$t_{\rm opt}$ that minimize the error
due to coupling imperfections in the $N=3$ case. The gate fidelity is defined in Eq. (\ref{eq:Fidelity}). The 
corresponding Hamiltonian is given in Eq. (\ref{eq:Himperfect}). Here, $\kappa = 0.05$, $\eta_{23}=0.02$, 
and $0\leq \eta_{13} \leq 0.10$. Fidelity curve for the original protocol run with no corrections is also shown.
}
\end{figure*}

Achieving identical couplings on all qubit pairs is experimentally difficult. One way to reduce errors  
due to coupling imperfections is to adjust the {\it final} qubit rotations and the entangling time in such a way 
as to maximize the fidelity of the resulting state. Figure \ref{fig:1} shows the result of numerical optimization 
with respect to the Frobenius distance between the generated state and the GHZ target,
\BEq
\label{eq:Fidelity}
{\cal F} = 1 - \sqrt{(\bra \psi|_{\rm opt} - \bra {\rm GHZ}|)(|\psi\ket_{\rm opt} - |{\rm GHZ}\ket)},
\EEq
for $N=3$. Here the Hamiltonian is given by
\BEq
\label{eq:Himperfect}
H_{\rm int} =  
(g_{12}/2)\left[
 \left(
 \sigma_x^1\sigma_x^2+\sigma_y^1\sigma_y^2\right)
+(1-\eta_{23})\left( \sigma_x^2\sigma_x^3+\sigma_y^2\sigma_y^3\right)
 +(1-\eta_{13})\left( \sigma_x^1\sigma_x^3+\sigma_y^1\sigma_y^3\right)
 +\kappa \left( \sigma_z^1\sigma_z^2  + \sigma_z^2\sigma_z^3 +  \sigma_z^1\sigma_z^3
 \right)
 \right],
\EEq
where $g_{12}>0$ is the reference coupling on qubits 1 and 2, $\eta_{23}, \eta_{13}\geq 0 $ are the coupling 
errors on pairs 23 and 13, and $\kappa$ is the $ZZ$ coupling, which for simplicity is assumed to be constant 
throughout 
the network (the results of this section do not depend on this assumption). Our simulations indicate that the 
best fidelity is achieved when the  initial pulse is chosen to 
generate the uniform superposition state, similar to how it is done in the ideal case with
$\eta_{23}=\eta_{13}=0$. A typical optimized state, here shown at $\kappa = 0.05$, 
$\eta_{23}=0.02$, and $\eta_{13}=0.06$, has the form
\BEq
|\psi\ket_{\rm opt} =
\begin{pmatrix}
  0.707099\cr
  0.000692\cr
 -0.001956\cr
  0.002566\cr
  0.002566\cr
 -0.001956\cr
  0.000692\cr
  0.707099
\end{pmatrix},
\EEq
which corresponds to $t_{\rm opt}/(\pi/2g_{12}(1-\kappa)) = 1.0505$, 
$\alpha_{\rm opt}^{1,2,3}/(\pi/2) = 0.9785,  0.9713,   0.9825$, and
${\cal F}=0.9953$. This may be compared to the non-corrected state,
\BEq
|\psi\ket_{\rm original} =
\begin{pmatrix}
0.706616\cr
  0.000697 + 0.015431i\cr
 -0.002792 + 0.010929i\cr
  0.002988 + 0.017844i\cr
  0.002988 + 0.017844i\cr
 -0.002792 + 0.010929i\cr
  0.000697 + 0.015431i\cr
  0.706616
\end{pmatrix},
\EEq
that has ${\cal F} = 0.9628$,
which we get by running the original protocol at the same values of $\kappa$ and
$\eta_{23}$, $\eta_{13}$.

We have also performed optimization in the $N=4$ case (not shown) where it was found that in order to 
maximize the fidelity of the resulting state only the entangling time $t_{\rm GHZ}$ and the final $Z$-rotation 
on the first qubit have to be adjusted. The $R_1$ pulse appearing on both sides of $U_{\rm ent}$ should, 
however, remain the same. 

The following argument indicates that similar optimization results may be expected for arbitrary $N$.
Assuming coupling errors in the interaction
Hamiltonian to be sufficiently small, their effect is to
perturb (in general all of) the $N+1$ eigenvalues $\lambda_j$ corresponding 
to the $|W_j\ket$ states.  If we then optimize the fidelity on the evolution
time $t_{\rm GHZ}$ and $N$ final single-qubit rotation angles, we have
$N+1$ parameters to do so. We thus have $N+1$ parameters in which to
correct $N+1$ state coefficient errors due to $N+1$ perturbed phases coming
from the perturbed eigenvalues. Minimizing the Frobenius distance
needed for the fidelity is essentially solving a nonlinear least
squares problem.  Under only weak conditions, we will have a
solution guaranteed as we have $N+1$ parameters at our disposal.
If there were errors in the initial fully uniform superposition state,
then the $N+1$ parameters would be insufficient in number to correct errors from {\em both} the 
initial state and a perturbed Hamiltonian. Improvements would be possible, but the correct
initial state would then seem to be the best place to start from when precisely $N+1$
parameters are available to later correct errors in the protocol.
This argument shows that finding corrected fidelity by adjusting the entangling time and the final 
$N$ rotation angles is {\it always} possible. It does not state how well the correction can be done, but 
the $N=3$ and $N=4$ cases presented above give illustrations.

\section{Conclusion}
In summary, we have demonstrated how the single-step GHZ protocol proposed for triangularly coupled 
superconducting qubit system can be extended to a fully connected network of an arbitrary number of qubits. 
Two types of protocols can be identified here, depending on whether $N$ is odd or even. When $N$ is odd, 
the protocol (up to an overall insignificant phase) is the same as in the three-qubit case; it is given by a 
generic sequence $X_{\pi/2}U_{\rm ent}Y_{\pi/2}|0\ket^{\otimes N}$, with $X_{\pi/2}$ and $Y_{\pi/2}$ 
representing the corresponding $x$ and $y$ rotations performed on the $N$ qubits. When $N$ is even, 
the final rotation gets modified and the sequence  becomes 
$Z^1_\theta Y_{\pi/2}U_{\rm ent}Y_{\pi/2}|0\ket^{\otimes N}$, where $Z^1_\theta$ is a $z$-rotation on 
the first qubit by angle $\theta$ which is given in Eq. (\ref{eq:theta}). As a subproblem to describe the 
dynamics in the $|W_j\ket $ subspaces, we exactly solved for the corresponding eigenenergies.

We emphasize again that implementing identical couplings on all pairs of qubits is a difficult task. Here we 
proposed a possible approach to reducing the errors due to coupling imperfections in the experimentally 
important low-dimensional cases.

\appendix
\section{Proof of the eigenvalue formula, Eq. (\ref{eq:evalues}).}
\label{sec:appendix1}

\subsection{Raising and lowering operators}

For the
purposes of this Appendix the matrix representations of states
$|0\ket=(0,1)^{\rm T}$ and $|1\ket=(1,0)^{\rm T}$ are taken.

We first define the raising and lowering operators $S_\pm^k=\frac{1}{2}\left(\sigma_x^k\pm i\sigma_y^k\right)$
with the following properties:
\begin{gather}
 S_-|0\rangle=0,\,\,S_+|1\rangle=0,\,\,(S_-)^2=(S_+)^2=0,\,\,S_+|0\rangle=|1\rangle,\,\,S_-|1\rangle=|0\rangle,
\nonumber\\
\sigma_z^kS_+^k=S_+^k,\,\,\sigma_z^kS_-^k=S_-^k,\,\,S_+^k\sigma_z^k=-S_+^k,\,\,S_-^k\sigma_z^k=S_-^k,
\nonumber\\
\sigma_z^k=\mathbf{I}_2-2S_-^kS_+^k,\,\,\sigma_z|0\rangle=-|0\rangle,\,\,\sigma_z|1\rangle=|1\rangle,
\end{gather}
and
\BEq
\left[S_+^{k_1},S_-^{k_2}\right]=\delta_{k_1k_2}\sigma_z^{k_2}, \quad
\left[\sigma_z^{k_1},S_\pm^{k_2}\right]=\pm2\delta_{k_1k_2}S_\pm^{k_2},
\EEq
It will be useful to define the operators
\begin{equation}
 \Sigma_\pm=\sum_{k=1}^NS_\pm^k,\,\,\Sigma_z=\sum_{k=1}^N\sigma_z^k,
\end{equation}
obeying
\begin{align}
 \left[\Sigma_+,\Sigma_-\right]=\Sigma_z, \quad
 \left[\Sigma_z,\Sigma_\pm\right]=\pm2\Sigma_\pm.
\end{align}
Using these commutation relations we arrive at the following useful products.

\textbf{Lemma 1:} 
\BEq
\Sigma_z(\Sigma_\pm)^j=(\Sigma_\pm)^j\Sigma_z\pm2j(\Sigma_\pm)^j.
\EEq

\textbf{Proof (by induction).} For $j=1$,
\[
 \left[\Sigma_z,\Sigma_\pm\right]=\Sigma_z\Sigma_\pm-\Sigma_\pm\Sigma_z=\pm2\Sigma_\pm, \quad 
 \Sigma_z\Sigma_\pm=\Sigma_\pm\Sigma_z\pm2\Sigma_\pm  .
\]
Assuming that $\Sigma_z(\Sigma_\pm)^n=(\Sigma_\pm)^n\Sigma_z\pm2n(\Sigma_\pm)^n$, we have
\begin{align}
 \Sigma_z(\Sigma_\pm)^{n+1}&=\left(\Sigma_z(\Sigma_\pm)^{n}\right)\Sigma_\pm \nonumber \\
 &=\left((\Sigma_\pm)^n\Sigma_z\pm2n(\Sigma_\pm)^n\right)\Sigma_\pm\nonumber\\
 &=(\Sigma_\pm)^n(\Sigma_\pm\Sigma_z\pm2\Sigma_\pm)\pm2n(\Sigma_\pm)^{n+1}\nonumber\\
 &=(\Sigma_\pm)^{n+1}\Sigma_z\pm2(n+1)(\Sigma_\pm)^{n+1}. 
 \quad \blacksquare \nonumber
\end{align}

\textbf{Lemma 2:} 
\BEq
\Sigma_-(\Sigma_+)^j=(\Sigma_+)^j\Sigma_--j(\Sigma_+)^{j-1}\Sigma_z-j(j-1)(\Sigma_+)^{j-1}.
\EEq

\textbf{Proof (by induction).}  For $j=1$,
\begin{align*}
 [\Sigma_+,\Sigma_-]=\Sigma_+\Sigma_--\Sigma_-\Sigma_+=\Sigma_z,
 \quad \Sigma_-\Sigma_+=\Sigma_+\Sigma_--\Sigma_z .
\end{align*}
For $j=2$,
\begin{align*}
 \Sigma_-\Sigma_+\Sigma_+&=(\Sigma_+\Sigma_--\Sigma_z)\Sigma_+\\
 &=\Sigma_+(\Sigma_+\Sigma_--\Sigma_z)-\Sigma_+\Sigma_z-2\Sigma_+\\
 &=(\Sigma_+)^2\Sigma_--2\Sigma_+\Sigma_z-2\Sigma_+ .
\end{align*}
Now, assuming $\Sigma_-(\Sigma_+)^n=(\Sigma_+)^n\Sigma_--n(\Sigma_+)^{n-1}\Sigma_z-n(n-1)(\Sigma_+)^{n-1}$, 
we get
\begin{align*}
 \Sigma_-(\Sigma_+)^{n+1}&=(\Sigma_-(\Sigma_+)^{n})\Sigma_+\\
 &=(\Sigma_+)^n(\Sigma_-\Sigma_+)-n(\Sigma_+)^{n-1}(\Sigma_z\Sigma_+)-n(n-1)(\Sigma_+)^{n}\\
 &=(\Sigma_+)^{n+1}\Sigma_--(\Sigma_+)^n\Sigma_z-n(\Sigma_+)^{n}\Sigma_z-2n(\Sigma_+)^n-n(n-1)(\Sigma_+)^{n}\\
 &=(\Sigma_+)^{n+1}\Sigma_--(n+1)(\Sigma_+)^{n}\Sigma_z-(n(n-1)+2n)(\Sigma_+)^{n}\\
 &=(\Sigma_+)^{n+1}\Sigma_--(n+1)(\Sigma_+)^{n}\Sigma_z-((n+1)((n+1)-1))(\Sigma_+)^{n}. \quad \blacksquare
\end{align*}

We now generate partial uniform superpositions (generalized $W$ states) using the raising and lowering operators:
\begin{align}
 |W_0\rangle&=1|0\rangle^{\otimes N},\nonumber\\
 |W_1\rangle&=\frac{1}{\sqrt{C^1_N}}\sum_{k=1}^N S_+^k|0\rangle^{\otimes N},\nonumber\\
 |W_2\rangle&=\frac{1}{\sqrt{C^2_N}}\sum_{k_1=1}^{N-1}\sum_{k_2>k_1}^{N} S_+^{k_1}S_+^{k_2}
 |0\rangle^{\otimes N},\nonumber\\
 &\vdots\nonumber\\
 |W_j\rangle&=\frac{1}{\sqrt{C^j_N}}\sum_{k_1=1}^{N-j+1}\sum_{k_2>k_1}^{N-j+2}...\sum_{k_{j-1}>k_{j-2}}^{N-1}
 \sum_{k_{j}>k_{j-1}}^{N}S_+^{k_1}S_+^{k_2}...S_+^{k_{j-1}}S_+^{k_{j}}|0\rangle^{\otimes N}.
\end{align}
If we take into account all double counting we can see that these states can also be written in a form
\begin{align}
 |W_0\rangle&=1|0\rangle^{\otimes N},\nonumber\\
 |W_1\rangle&=\frac{1}{\sqrt{C^1_N}}\sum_{k=1}^N S_+^k|0\rangle^{\otimes N}=\frac{1}{\sqrt{C^1_N}}
 \Sigma_+|0\rangle^{\otimes N},\nonumber\\
 |W_2\rangle&=\frac{1}{2!\sqrt{C^2_N}}\sum_{k_1=1}^{N}\sum_{k_2=1}^{N} S_+^{k_1}S_+^{k_2}|0\rangle^{\otimes N}
 =\frac{1}{2!\sqrt{C^2_N}}(\Sigma_+)^2|0\rangle^{\otimes N},\nonumber\\
 &\vdots\nonumber\\
 |W_j\rangle&=\frac{1}{j!\sqrt{C^j_N}}\sum_{k_1=1}^{N}\sum_{k_2=1}^{N}...\sum_{k_{j-1}=1}^{N}
 \sum_{k_{j}=1}^{N}S_+^{k_1}S_+^{k_2}...S_+^{k_{j-1}}S_+^{k_{j}}|0\rangle^{\otimes N}=\frac{1}{j!\sqrt{C^j_N}}
 (\Sigma_+)^{j}|0\rangle^{\otimes N}.
\end{align}
Thus,
\begin{align}
 \Sigma_+|W_j\rangle&=\frac{1}{j!\sqrt{C^j_N}}(\Sigma_+)^{j+1}|0\rangle^{\otimes N},\nonumber\\
 &=\frac{1}{j!\sqrt{C^j_N}}\left((j+1)!\sqrt{C^{j+1}_N}\right)|W_{j+1}\rangle,\nonumber\\
 &=\sqrt{(N-j)(j+1)}|W_{j+1}\rangle .
\end{align}
Using the lemmas and the facts that $\Sigma_-|0\rangle^{\otimes N}=0$ and $\Sigma_z|0\rangle^{\otimes N}
=-N|0\rangle^{\otimes N}$, we find:
\begin{align}
 \Sigma_-|W_j\rangle&=\frac{1}{j!\sqrt{C^j_N}}(\Sigma_-(\Sigma_+)^j)|0\rangle^{\otimes N},\nonumber\\
 &=\frac{1}{j!\sqrt{C^j_N}}((\Sigma_+)^j\Sigma_--j(\Sigma_+)^{j-1}\Sigma_z-j(j-1)(\Sigma_+)^{j-1})
 |0\rangle^{\otimes N},\nonumber\\
 &=\frac{j(N-j+1)}{j!\sqrt{C^j_N}}(\Sigma_+)^{j-1}|0\rangle^{\otimes N},\nonumber\\
 &=\sqrt{j(N-j+1)}|W_{j-1}\rangle,
\end{align}
and
\begin{align}
 \Sigma_z|W_j\rangle&=\frac{1}{j!\sqrt{C^j_N}}\Sigma_z(\Sigma_+)^{j}|0\rangle^{\otimes N},\nonumber\\
 &=\frac{1}{j!\sqrt{C^j_N}}((\Sigma_+)^j\Sigma_z+2j(\Sigma_+)^j)|0\rangle^{\otimes N},\nonumber\\
 &=\frac{1}{j!\sqrt{C^j_N}}(-N+2j)(\Sigma_+)^j|0\rangle^{\otimes N},\nonumber\\
 &=(2j-N)|W_j\rangle .
\end{align}

\subsection{Eigenvalues of $H_{\rm int}$}

First notice that 
\begin{align}
 \sigma_x^{k_1}&\sigma_x^{k_2}+\sigma_y^{k_1}\sigma_y^{k_2}
=2\left(S_+^{k_1}S_-^{k_2}+S_-^{k_1}S_+^{k_2}\right),\nonumber
\end{align}
from which it follows that
\begin{equation}
 H_{\rm int}=\frac{1}{2}\sum_{k_1=1}^{N-1}\sum_{k_2=k_1+1}^N\left(2g\left(S_+^{k_1}S_-^{k_2}+S_-^{k_1}
 S_+^{k_2}\right)+g_z\sigma_z^{k_1}\sigma_z^{k_2}\right).
\end{equation}
Also,
\begin{equation*}
 \Sigma_z\Sigma_z=\sum^N_{k_1,k_2=1}\sigma_z^{k_1}\sigma_z^{k_2}=\sum^N_{k_1\neq k_2=1}
 \sigma_z^{k_1}\sigma_z^{k_2}+\sum^N_{k=1}(\sigma_z^{k})^2=2\sum^{N-1}_{k_1=1}
 \sum^{N}_{k_2=k_1+1}\sigma_z^{k_1}\sigma_z^{k_2}+N\mathbf{I},
\end{equation*}
and similarly,
\begin{gather*}
 \Sigma_+\Sigma_-=\sum^N_{k_1,k_2=1}S_+^{k_1}S_-^{k_2}=\sum^N_{k_1\neq k_2=1}S_+^{k_1}S_-^{k_2}
 +\sum^N_{k=1}S_+^{k}S_-^{k}=2\sum^{N-1}_{k_1=1}\sum^{N}_{k_2=k_1+1}S_+^{k_1}S_-^{k_2}
 +\sum^N_{k=1}S_+^{k}S_-^{k},\\
 \Sigma_-\Sigma_+=\sum^N_{k_1,k_2=1}S_-^{k_1}S_+^{k_2}=\sum^N_{k_1\neq k_2=1}S_-^{k_1}S_+^{k_2}
 +\sum^N_{k=1}S_-^{k}S_+^{k}=2\sum^{N-1}_{k_1=1}\sum^{N}_{k_2=k_1+1}S_-^{k_1}S_+^{k_2}
 +\sum^N_{k=1}S_-^{k}S_+^{k}.
\end{gather*}
Using the fact that $\sum^N_{k=1}S_+^{k}S_-^{k}+\sum^N_{k=1}S_-^{k}S_+^{k}=N\mathbf{I}$, we re-write 
$H_{\rm int}$ in the form
\begin{align}
 H_{\rm int}&=\frac{1}{2}\left(2g\sum_{k_1=1}^{N-1}\sum_{k_2=k_1+1}^NS_+^{k_1}S_-^{k_2}
 +2g\sum_{k_1=1}^{N-1}\sum_{k_2=k_1+1}^NS_+^{k_2}S_-^{k_1}+\tilde{g}\sum_{k_1=1}^{N-1}
 \sum_{k_2=k_1+1}^N\sigma_z^{k_1}\tilde{g}^{k_2}\right)\nonumber\\
 &=\frac{1}{2}\left(2g\left(\frac{1}{2}(\Sigma_+\Sigma_--\sum^N_{k=1}S_+^{k}S_-^{k})\right)
 +2g\left(\frac{1}{2}(\Sigma_-\Sigma_+-\sum^N_{k=1}S_-^{k}S_+^{k})\right)
 +\tilde{g}\left(\frac{1}{2}(\Sigma_z\Sigma_z-N\mathbf{I})\right)\right),\nonumber\\
 &=\frac{1}{4}\left(2g\left(\Sigma_+\Sigma_-+\Sigma_-\Sigma_+\right)
 +\tilde{g}\Sigma_z\Sigma_z-(2g+\tilde{g})N\mathbf{I}\right),
\end{align}
or
\BEq
 H_{\rm int}=\frac{1}{4}\left(2gH_{\rm int}^g+\tilde{g}H_{\rm int}^{\tilde g}\right),
 \EEq
where
\BEq
 H_{\rm int}^g=\Sigma_+\Sigma_-+\Sigma_-\Sigma_+-N\mathbf{I},
\quad  H_{\rm int}^{\tilde g}=\Sigma_z\Sigma_z-N\mathbf{I}.
 \EEq
We now act with $H_{\rm int}^g$ on the partial uniform states, 
\BEqA
 H_{\rm int}^g|W_j\rangle&=&\left(\Sigma_+\Sigma_-+\Sigma_-\Sigma_+-N\mathbf{I}\right)
 |W_j\rangle,\nonumber\\
&=&\sqrt{j(N-j+1)}\Sigma_+|W_{j-1}\rangle+\sqrt{(N-j)(j+1)}\Sigma_-|W_{j+1}\rangle-N
|W_j\rangle,\nonumber\\
&=&\left(j(N-j+1)+(N-j)(j+1)-N\right)|W_j\rangle,\nonumber\\
&=&2j(N-j)|W_j\rangle,
\EEqA
and similarly,
\BEq
 H_{\rm int}^{\tilde g}|W_j\rangle=(\Sigma_z\Sigma_z-N\mathbf{I})|W_j\rangle\\
 =((2j-N)^2-N)|W_j\rangle
\EEq
It is then easy to see that $H_{\rm int}|W_j\rangle=\lambda_j(N)|W_j\rangle$, where
\BEq
 \lambda_j(N)=\frac{1}{4}\left(4j(N-j)g+((2j-N)^2-N)\tilde{g}\right)
 =j(N-j)(g-\tilde{g})+C^2_N\left({\tilde g}/2\right).
\EEq
This completes the proof of the eigenvalue formula.

\section{Proof of Eq. (\ref{eq:2})}
\label{sec:appendix2}

Here, $N$ is odd. Using the fact that for $y \in\{0,1\}$,
\BEq
e^{-i \pi \sigma_x/4} |{y}\ket={{(-i)^y} \over \sqrt{2}}[|{0}\ket-i(-1)^y|{1}\ket]
={{(-i)^y} \over \sqrt{2}}\sum_{z \in \{0,1\}} (-i)^z(-1)^{yz} |{z}\ket,
\EEq
we see that the action of $R_2^{-1}$ on arbitrary direct-product states is given by
\BEq
R_2^{-1}|y_1 y_2 \ldots y_N\ket={{i^{y_1+y_2+\ldots+y_N}} \over 2^{N/2}} 
\sum_{z_k \in \{0,1\}} i^{z_1+z_2+\ldots+z_N}(-1)^{y_1z_1+y_2z_2+\ldots+y_Nz_N} 
|{z_1}\ket |{z_{2}}\ket \ldots |{z_N}\ket .
\EEq
We now let $z_{tot} = \sum_{k=1}^N z_k$ when occurring in such sums, and write
more compactly
\BEq
R_2^{-1}|y_1 y_2 \ldots y_N\ket={{i^{y_1+y_2+\ldots+y_N}} \over 2^{N/2}} 
\sum_{z_k \in \{0,1\}} i^{z_{tot}}(-1)^{y \cdot z}|{z_1}\ket |{z_{2}}\ket\ldots |{z_N}\ket.
\EEq
We note that $0 \leq z_{\rm tot} \leq N$. We then have
\BEq
R_2^{-1}|0\ket^{\otimes N}={1 \over 2^{N/2}} \sum_{z_k} i^{z_{tot}} 
|{z_1}\ket \ldots |{z_N}\ket,
\quad 
R_2^{-1}|{1}\ket^{\otimes N}={i^N \over 2^{N/2}} \sum_{z_j} i^{z_{tot}} (-1)^{z_{tot}} 
|{z_1}\ket \ldots |{z_N}\ket. 
\EEq
Therefore,
\BEq
R_2^{-1} |{{\rm {GHZ}}}\ket_{{\rm {N ~odd}}}=R_2^{-1} {1 \over \sqrt{2}} 
(|{0}\ket^{\otimes N}+|{1}\ket^{\otimes N})
={1 \over {\sqrt{2}2^{N/2}}}\sum_{z_j} i^{z_{tot}}[1+(-1)^{z_{tot}}i^N]
|{z_1}\ket \ldots |{z_N}\ket.  
\EEq
Replacing the sum over $z_k$'s by $z_{tot}$, which is just an integer $j$
in the range $0,\ldots,N$, gives
\BEq
R_2^{-1} |{{\rm GHZ}}\ket_{{\rm N ~odd}}={1 \over {2^{N/2}}}\sum_{j=0}^N 
\sqrt{C_N^j} \frac{i^j  [1+(-1)^j i^N]}{\sqrt{2}}|{W_j}\ket. 
\EEq

\section{Proof of Eq. (\ref{eq:3})}
\label{sec:appendix3}

Here, $N$ is even. Note that for $x \in\{0,1\}$,
\BEq
e^{-i \pi \sigma_y/4} |{x}\ket={1 \over \sqrt{2}}[(-1)^x|{0}\ket+|{1}\ket]
={1 \over \sqrt{2}}\sum_{z \in \{0,1\}} (-1)^{x(z-1)} |{z}\ket,
\EEq
which gives
\BEq
R_1^{-1}|{x}\ket={1 \over 2^{N/2}}\sum_{z_j \in \{0,1\}} (-1)^{x\cdot z} (-1)^
{z_{tot}} |{z_1}\ket |{z_2}\ket \ldots |{z_N}\ket.
\EEq
Since
\BEq
R_1^{-1}|{0}\ket^{\otimes N}={1 \over 2^{N/2}} \sum_{z_j} (-1)^{z_{tot}} 
|{z_1}\ket \ldots |{z_N}\ket,
\quad 
R_1^{-1}|{1}\ket^{\otimes N}={1 \over 2^{N/2}} \sum_{z_j} |{z_1}\ket \ldots 
|{z_N}\ket,
\EEq
we have
\BEqA
R_1^{-1} [R_{3}]^{-1}|{{\rm {GHZ}}}\ket_{{\rm {N ~even}}}
&=& R_1^{-1} {1 \over \sqrt{2}} (|{0}\ket^{\otimes N}+e^{i\varphi}|{1}\ket^{\otimes N})
\nonumber \\
&=&{1 \over {\sqrt{2}2^{N/2}}}\sum_{z_j}[(-1)^{z_{tot}}+e^{i\varphi}]\sum_{z_j} 
|{z_1}\ket \ldots |{z_N}\ket \nonumber \\
&=&{1 \over {2^{N/2}}}\sum_{j=0}^N \sqrt{C_N^j} \frac{[(-1)^j + e^{i\varphi}]}{\sqrt{2}} |{W_j}\ket,
\EEqA
where $\varphi=-\theta(N)$.

\begin{acknowledgments}

The work of AG was supported by IARPA 
under grant W911NF-08-1-0336 and by the NSF under grant CMS- 
0404031. The work of MWC and RD was supported in 
part by a grant from Northrop Grumman. We thank Sahel Ashhab, John Martinis, and 
Matthew Neeley for useful discussions.

\end{acknowledgments}

\end{document}